\documentclass[fleqn,usenatbib]{mnras}

\usepackage{newtxtext,newtxmath}
\usepackage[T1]{fontenc}

\DeclareRobustCommand{\VAN}[3]{#2}
\let\VANthebibliography\thebibliography
\def\thebibliography{\DeclareRobustCommand{\VAN}[3]{##3}\VANthebibliography}

\usepackage{graphicx}	% Including figure files
\usepackage{amsmath}	% Advanced maths commands
\usepackage{subfigure}
\usepackage{float}
\usepackage{bm}
\usepackage{orcidlink}
\usepackage{hyperref}
\newcommand{\mpchi}{\,{\rm {cMpc/h}}}

\newcommand{\msunh}{{\rm M}_{\sun}/h}

\def\thetalop{\,{\rm \theta_{lop}}}

\title[Lopsided satellite distribution in MTNG]{The origin of lopsided satellite galaxy distribution around isolated systems in MillenniumTNG}

\author[Liu et al.]{%
Yikai Liu$^{1,2}$\orcidlink{0009-0007-2848-7454},
Peng Wang$^{1,3}$\thanks{E-mail: pwang@shao.ac.cn}\orcidlink{0000-0003-2504-3835},
Hong Guo$^{1}$\thanks{E-mail: guohong@shao.ac.cn}\orcidlink{0000-0003-4936-8247}, 
Volker Springel$^4$\orcidlink{0000-0001-5976-4599},
Sownak Bose$^5$\orcidlink{0000-0002-0974-5266},
R\"udiger Pakmor$^4$\orcidlink{0000-0003-3308-2420},
\newauthor
Lars Hernquist$^6$
\vspace*{6pt} \\%
$^{1}$Shanghai Astronomical Observatory, Chinese Academy of Sciences, Shanghai 200030, China\\%
$^{2}$University of Chinese Academy of Sciences, Beijing 100049, China\\%
$^{3}$Astronomical Research Center, Shanghai Science \& Technology Museum, Shanghai, 201306, China\\%
$^{4}$Max-Planck-Institut f\"ur Astrophysik, Karl-Schwarzschild-Str. 1, D-85748 Garching, Germany\\%
$^{5}$Institute for Computational Cosmology, Department of Physics, Durham University, South Road, Durham DH1 3LE, UK\\%
$^{6}$Harvard-Smithsonian Center for Astrophysics, 60 Garden St, Cambridge, MA 02138, USA\\%
}

\begin{document}
\label{firstpage}
\pagerange{\pageref{firstpage}--\pageref{lastpage}}
\maketitle

%%%%%%%%%%%%%%%%%%%%%%%%%%%%%%%%%%%%%%%%%%%%%%%%%%%%%
\begin{abstract}
Dwarf satellites in galaxy groups are distributed in an anisotropic and asymmetric manner, which is called the ``lopsided satellite distribution''. This lopsided signal has been observed not only in galaxy pairs but also in isolated systems. However, the physical origin of the lopsided signal in isolated systems is still unknown. In this work, we investigate this in the state-of-the-art hydrodynamical simulation of the MillenniumTNG Project by tracing each system back to high redshift. We find that the lopsided signal is dominated by satellites located in the outer regions of the halo and is also dominated by recently accreted satellites. The lopsided signal originates from the anisotropic accretion of galaxies from the surrounding large-scale structure and that, after accretion, the nonlinear evolution of satellites inside the dark-matter halo weakens the lopsidedness. The signal decreases as cosmic time passes because of a competition between anisotropic accretion and internal evolution within dark matter halos. Our findings provide a useful perspective for the study of galaxy evolution, especially for the origin of the spatial satellite galaxy distributions.
\end{abstract}
%%%%%%%%%%%%%%%%%%%%%%%%%%%%%%%%%%%%%%%%%%%%%%%%%%%%%

%%%%%%%%%%%%%%%%%%%%%%%%%%%%%%%%%%%%%%%%%%%%%%%%%%%%%
%%%%%%%%%%
%   Introduction
%%%%%%%%%%
\section{Introduction}\label{sec:intro}
The matter distribution across the Universe is homogeneous and isotropic on large scales, which is in agreement with the predictions of the $\Lambda$CDM cosmological model. However, the anisotropy and inhomogeneity of the matter distribution increases in localised regions. On scales of a few to hundreds of megaparsecs, the distribution of matter is characterised by a web-like structure known as the cosmic web, which is made up of knots, filaments, sheets, and voids \citep{1970A&A.....5...84Z, 1991ApJ...379..440B}. Within galaxy groups, the phase-space distribution of dwarf satellite galaxies around their central galaxies is anisotropic \citep{1997ApJ...478L..53Z,2005ApJ...629..219Z,2005ApJ...628L.101B}.

In recent decades, numerous studies have been conducted on the phase-space distribution of satellites \citep{2004ApJ...603....7K,2005MNRAS.364..424W,2008MNRAS.386L..52K,2011MNRAS.411.1525L,2021Galax...9...66P}. After a long period of debate \citep{1968PASP...80..252S,1969ArA.....5..305H}, it is now widely accepted that satellites are preferentially located around the main axis of their central galaxies \citep{2005ApJ...628L.101B,2005MNRAS.363..146L,2010ApJ...709.1321A,2013ApJ...768...20L,2018A&A...613A...4W,2019MNRAS.484.4325W,2020ApJ...893...87T}. This alignment signal depends on the colour of the galaxy. The red satellites around the red centrals show the most significant alignment, while the blue satellites around the blue centrals are more randomly distributed \citep{2006MNRAS.369.1293Y,2007MNRAS.376L..43A,2007MNRAS.378.1531K}. However, another issue of satellite phase-space distribution, the so-called ``satellite plane problem'', is still not well understood and controversially discussed \citep{2013MNRAS.436.2096S,2015ApJ...800...34G,2017ApJ...843...62M,2018MPLA...3330004P,2021NatAs...5.1185P,2023NatAs...7..481S,2023MNRAS.520.3937P}. In the Local Universe, the brightest satellites of the Milky Way \citep{2005A&A...431..517K, 2008ApJ...680..287M,2013MNRAS.435.2116P} and Andromeda \citep{2013Natur.493...62I,2013ApJ...766..120C,2020MNRAS.492..456W} are found to be located in a thin corotating plane, whereas such a satellite plane is rare \citep{2014ApJ...784L...6I} in cosmological simulations. Recently, a similar satellite plane was also discovered around the central galaxy Centaurus A \citep{2015ApJ...802L..25T,2018Sci...359..534M}.

In the Local Group, the Milky Way and M31 are considered as a galaxy pair, and there are 25 satellites located in the area between them. By using galaxy pairs taken from SDSS DR10 \citep{2000AJ....120.1579Y,2014ApJS..211...17A}, \cite{2016ApJ...830..121L} discovered that 8\% more satellites than would be expected from a uniform distribution are situated in the region between two hosts of a galaxy pair similar to the MW-M31 pair, which is known as the problem of the ``lopsided satellite distribution'' (LSD). Following these observations, \cite{2017ApJ...850..132P} identified a comparable signal of LSD in a galaxy pair in a $\Lambda$CDM cosmological N-body simulation.

Examining isolated systems such as the Milky Way (MW) and M31, it has been observed that 7 of the 11 brightest MW satellites and 21 of the 27 M31 satellites are located in the same hemisphere. Studies of other MW-like galaxies in the Local Volume, such as M101, have revealed that 7 of the 8 brightest satellites are located in one hemisphere \citep{2014ApJ...787L..37M,2019ApJ...885..153B,2020ApJ...893L...9B}. \cite{2020ApJ...898L..15B} investigated the satellite distribution around isolated galaxies in the NASA-Sloan Atlas catalogue and found that the probability distribution for the polar angles of the satellites differs significantly from a random distribution. \cite{2021ApJ...914...78W} further found that the lopsided signal depends on the mass, colour, and large-scale environment, with satellites that reside in the large radius of low-mass blue hosts exhibiting the most lopsided signal. \cite{2023ApJ...947...56S} also studied the lopsided satellite distribution by selecting isolated host galaxies and their satellites from mock redshift surveys of an N-body simulation and obtained results similar to those of \cite{2020ApJ...898L..15B}.

Observations and simulations have both demonstrated that the lopsided distribution of satellites is not exclusive to galaxy pairs, but also applies to isolated galaxies. The magnitude of the lopsidedness depends on both the mass and the colour. The physical source of LSD has been explored, with \cite{2014ApJ...793L..42B} suggesting that the lopsidedness of M31 is more likely due to a dwarf association falling into its halo than the tidal field of the Milky Way. \cite{2019MNRAS.488.3100G} found that the lopsided signal is generated from the first satellite approach and weakens over time due to interactions between satellite and central pairs. However, the origin of LSD in isolated systems remains a mystery. It is worth investigating whether the origins of LSD in isolated systems are similar to or completely different from those in galaxy pair systems, given the unique environment of isolated systems.

To understand the source of LSD in isolated systems, we investigate the history of these systems and study the development of the LSD signal in the state-of-the-art MilleniumTNG hydrodynamic simulation (see Section~\ref{sec:method} and the references therein for more details). The MilleniumTNG simulation is based on the IllustrisTNG \citep{2018MNRAS.475..676S, 2018MNRAS.480.5113M, 2018MNRAS.477.1206N, 2018MNRAS.475..624N, 2018MNRAS.473.4077P, 2018MNRAS.475..648P, 2017MNRAS.465.3291W} physics model but features a much larger volume at still high resolution, making it possible to obtain a substantial galaxy group sample to statistically explore the origin of LSD. 

Our paper is structured as follows. Section~\ref{sec:method} describes the simulation we use and how we measure the intensity of the lopsided signal. In Section~\ref{sec:result}, we look back at the evolution of the lopsided signal and try to identify its origin by comparing lopsided and nonlopsided systems in both the dark-matter halo and the satellites themselves. Section~\ref{sec:sum_dis} provides a summary and a discussion.

%%%%%%%%%%%%%%%%%%%%%%%%%%%%%%%%%%%%%%%%%%%%%%%%%%%%%
%%%%%%%%%%
%   Method
%%%%%%%%%%
\section{Data and Methodology}
\label{sec:method}

 \begin{figure}
\includegraphics[width=\columnwidth]{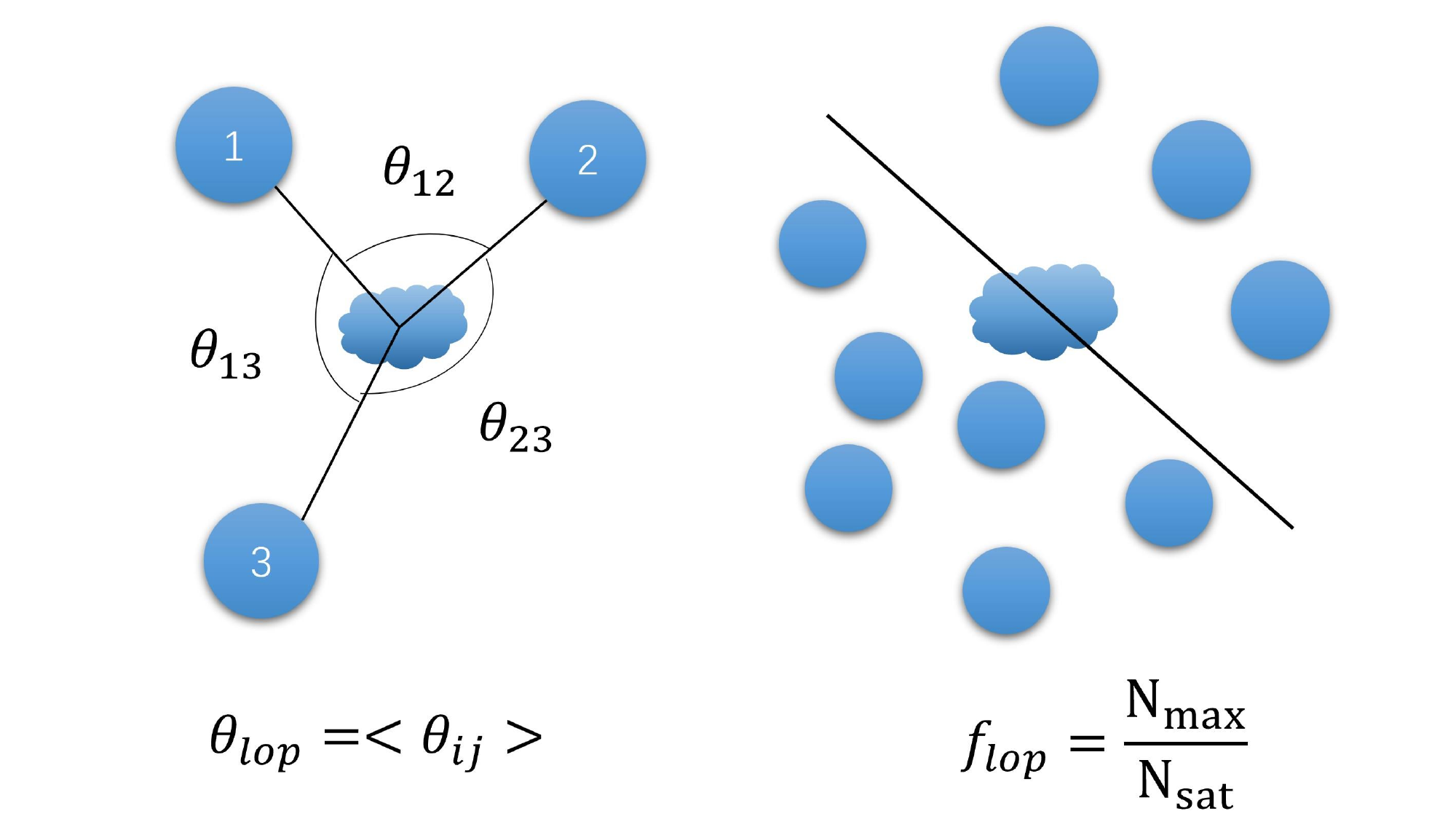}
    \caption{A schematic diagram illustrating the calculation of the lopsided angle $\theta_{\rm lop}$ (left) and lopsided fraction $f_{\rm lop}$ (right). The central galaxies are shown with cloud-like symbols located in the centre, and small circles indicate satellite positions around the central galaxies. In the left pane, $\rm \theta_{ij}$ is the angle between satellite pairs $i$ and $j$. In the right panel, $\rm N_{sat}$ is the total number of satellites in a given system, and $\rm N_{max}$ is the satellite number with the semicircle that contains the most satellite galaxies.}
\label{fig:def}
\end{figure}

We used the galaxy catalogue from the MTNG740 simulation of the MillenniumTNG Project \citep[hereafter MTNG;][]{2023MNRAS.524.2556H,2023MNRAS.524.2539P}. This simulation was carried out in a box size of $\rm (500 \ cMpc/h)^3$ or $(\rm  738.1 \ cMpc)^3$  with cosmological parameters taken from Planck \citep{2016A&A...594A..13P}, i.e., $\Omega_{m}=0.3089$, $\Omega_{\Lambda}=0.6911$, $\Omega_{b}=0.0486$ and $h=0.6774$. It contains $4320^3$ dark matter particles and $4320^3$ initial gas cells, with mass resolutions of $1.7\times 10^8\,{\rm M_{\odot}}$ and $3.1\times 10^7\,{\rm M_{\odot}}$, respectively. Detailed information on the MTNG project refer to papers of \cite{2023MNRAS.524.2556H,2023MNRAS.524.2539P}. More introductory papers of MTNG includes past lightcones \citep{2023MNRAS.tmp.2586B}, large-scale clustering \citep{2023MNRAS.524.2579B}, cosmological parameters analysis \citep{2023MNRAS.524.2489C}, intrinsic alignments of galaxies and haloes \citep{2023MNRAS.523.5899D}, weak lensing \citep{2023MNRAS.524.5591F}, halo occupation models (HODs) \citep{2023MNRAS.524.2507H}, satellite population model \citep{2023MNRAS.524.2524H} and galaxy population at high redshift \citep{2023MNRAS.524.2594K}.

Dark matter halos are identified by the Friend-of-Friend(FoF) algorithm and substructures are identified by the SUBFIND-HBT algorithm \citep{2021MNRAS.506.2871S}. The GADGET-4 code \citep{2021MNRAS.506.2871S} was used to construct the halo merger trees. The merger trees are self-contained, meaning that all subhalos are in the same tree as their progenitors and descendants, allowing us to easily trace the history of galaxy groups.

In this study, we selected central galaxies with halo viral mass $M_{200}$ greater than $10^{11}\msunh$ and at least two satellites with a total mass in their subhalos greater than $10^{10}\msunh$. We use the isolation criteria of \cite{2021ApJ...914...78W} to identify our target groups as isolated groups, which are groups whose distance from the nearest massive group (i.e. $M_{200}$ $\textgreater$ $10^{11}\msunh$) is greater than 2$\mpchi$. 
We note that two primary parameters determine the sample selection process: the mass of satellites and the distance used to define isolated systems. \cite{2021ApJ...914...78W} has studied a variety of combinations of these two parameters and found a weak dependence on them. For further information, we refer the reader to Figure 4 in \cite{2021ApJ...914...78W}. In the end, 35579 groups and 110346 satellites were selected for the following analysis.

\begin{figure*}
    \centering
    \includegraphics[scale=0.5]{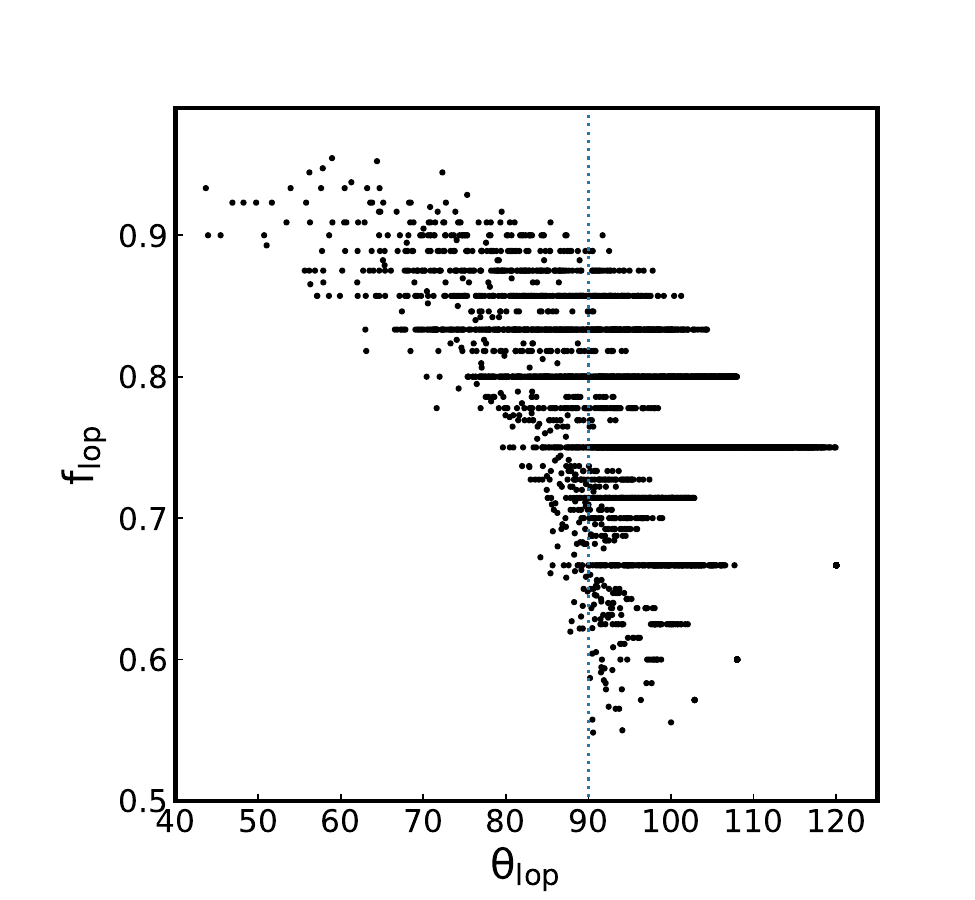}
    \hspace{2pt}
    \includegraphics[scale=0.5]{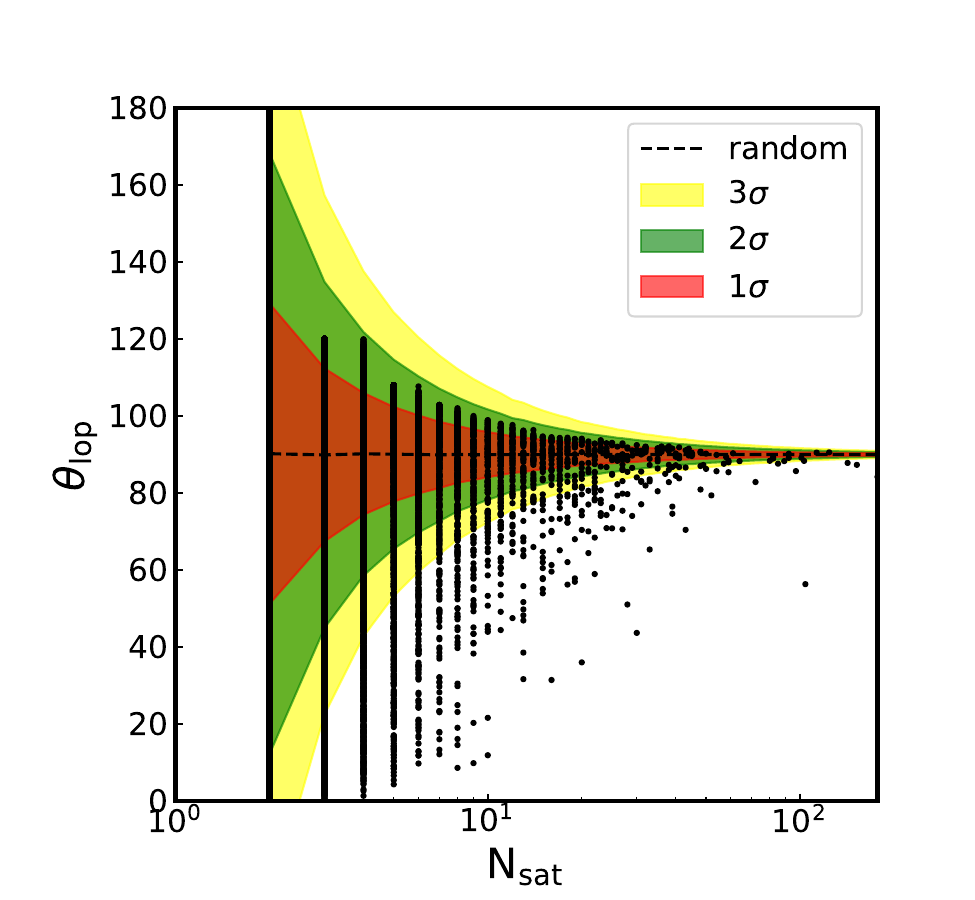}
    \caption{Left panel: Relation between the angle of lopsidedness and the fraction of lopsidedness. Each black dot represents a galaxy group in MTNG. The blue dashed vertical line is the lopsided angle expected for a random distribution, that is, 90$^{\circ}$. Right panel: lopsided signal $\rm \theta_{lop}$ as a function of the number of satellites per galaxy group. Shaded regions with different colours show random distributions from 1$\sigma$ to 3$\sigma$, as labelled in the legend. The black dots show the galaxy systems in MTNG.
    }
    \label{fig:lopsided}
\end{figure*}

To quantify the lopsidedness of a given isolated system, we define the lopsided angle, $\theta_{\rm lop}$, and the lopsided fraction $f_{\rm lop}$, as in Figure~\ref{fig:def}. To calculate $\rm \theta_{lop}$, we project all the satellite galaxies onto the $xy$ plane and place the central galaxy at the origin. As mentioned in \cite{2021ApJ...914...78W}, the viewing angle (that is, the choice of projection plane) does have an effect on the lopsided angle for a given system, but there is a strong linear correlation between the calculated lopsided signal projected in the 2D and 3D space. Thus, an arbitrarily chosen projection of the $xy$-plane is suitable and does not affect the statistical results. We calculate the lopsided angle as the average of angles between central-satellite galaxy pairs $\theta_{ij}$, i.e., 
\begin{equation} 
\theta_{\rm lop}=\left<\theta_{ij}\right>,
\end{equation} 
where the subscript $ij$ refers to satellite galaxies $i$ and $j$. The expected $\theta_{\rm lop}$ for a random distribution is close to $90^{\circ}$. However, in a lopsided system, satellite galaxies are usually situated on one side of the central, so the expected $\theta_{\rm lop}$ would be less than $90^{\circ}$. For example, the lopsided angle is $\theta_{\rm lop}=77.56^{\circ}$ for the MW.
%and $\theta_{\rm lop}=83.75^{\circ}$ for M101. 
It is important to remember that the lopsidedness of two systems cannot be compared simply by looking at their $\theta_{\rm lop}$. For galaxy groups with only two satellites, the $\theta_{\rm lop}$ for random distributions can vary greatly. On the other hand, the dispersion for galaxy groups with a large number of satellites is very small. Thus, the same $\theta_{\rm lop}$ can signify different levels of lopsidedness depending on the number of satellites.

The second quantity is the lopsided fraction, $f_{\rm lop}$. 
To calculate $f_{\rm lop}$, it is necessary to draw a line through the centre of the galaxy to divide the projected plane into two sections, with one of them containing the maximum number of satellites.
For a group with $N_{\rm sat}$ satellite galaxies, we can fit a straight line passing the centrals with the most satellites ($N_{\rm max}$) on one side in the $xy$-plane. Then the lopsided fraction is defined as
\begin{equation}
    f_{\rm lop} = \frac{N_{\rm max}}{N_{\rm sat}}.
\end{equation}
In the local universe, for example, the lopsided fraction is $f_{\rm lop} = 0.82$ for MW, 
%and $f_{\rm lop} = 0.77$ for M31. For M101, the lopsided fraction is $f_{\rm lop} = 0.87$ 
which shows an obvious lopsided signal of its satellite distribution.

The correlation between lopsided angles and lopsided fractions is demonstrated in Figure~\ref{fig:lopsided}. This correlation is particularly evident in a large number of satellite galaxies.  When the number of satellites is high, a larger $f_{\rm lop}$ implies a more pronounced lopsided signal. However, this is not the case for fewer satellites, as all $N_{\rm sat}=2$ groups have a $f_{\rm lop}$ of 1. Further details can be found in the appendix of \cite{2021ApJ...914...78W}.

The significance of a given system depends on the number of satellites, $N_{\rm sat}$. To determine whether a lopsided signal is inconsistent with a random distribution, we calculate the significance of $\theta_{\rm lop}$. To do this, for a system with a given $N_{\rm sat}$, we keep the distance between each satellite galaxy and the central galaxy fixed but randomise their angular distributions in the projected plane. We then repeat this randomisation 10,000 times and save the corresponding $\theta_{\rm lop, Ran}$ each time. We can then estimate the mean lopsided angle $\langle\theta_{\rm lop, Ran}\rangle$ for the random process and the corresponding $\sigma(\theta_{\rm lop, Ran})$ for each system with the given $N_{\rm sat}$. The significance is then calculated as
\begin{equation}
\label{equ:sig}
    \rm{significance} = \frac{\left|\theta_{\rm lop}-
    \langle\theta_{\rm lop, Ran}\rangle\right|}{\sigma(\theta_{\rm lop, Ran})}.
\end{equation}

The right panel of Figure~\ref{fig:lopsided} shows three different coloured shadows that signify 1$\sigma$, 2$\sigma$, and 3$\sigma$ of the lopsided angle of the random trials. Groups with different numbers of satellites show different levels of significance. A considerable number of systems ($\sim$ 5\% of the total) are located beyond 3$\sigma$, which demonstrates a clear lopsided signal. The findings in Figure~\ref{fig:lopsided} agree with those of \cite{2021ApJ...914...78W}.

%%%%%%%%%%%%% 
%   Result
%%%%%%%%%%%%% 
\section{Results}
\label{sec:result}
\subsection{General evolution of the lopsided signal}
    \label{subsec:evolution}

 \begin{figure*}
\includegraphics[width=\textwidth]{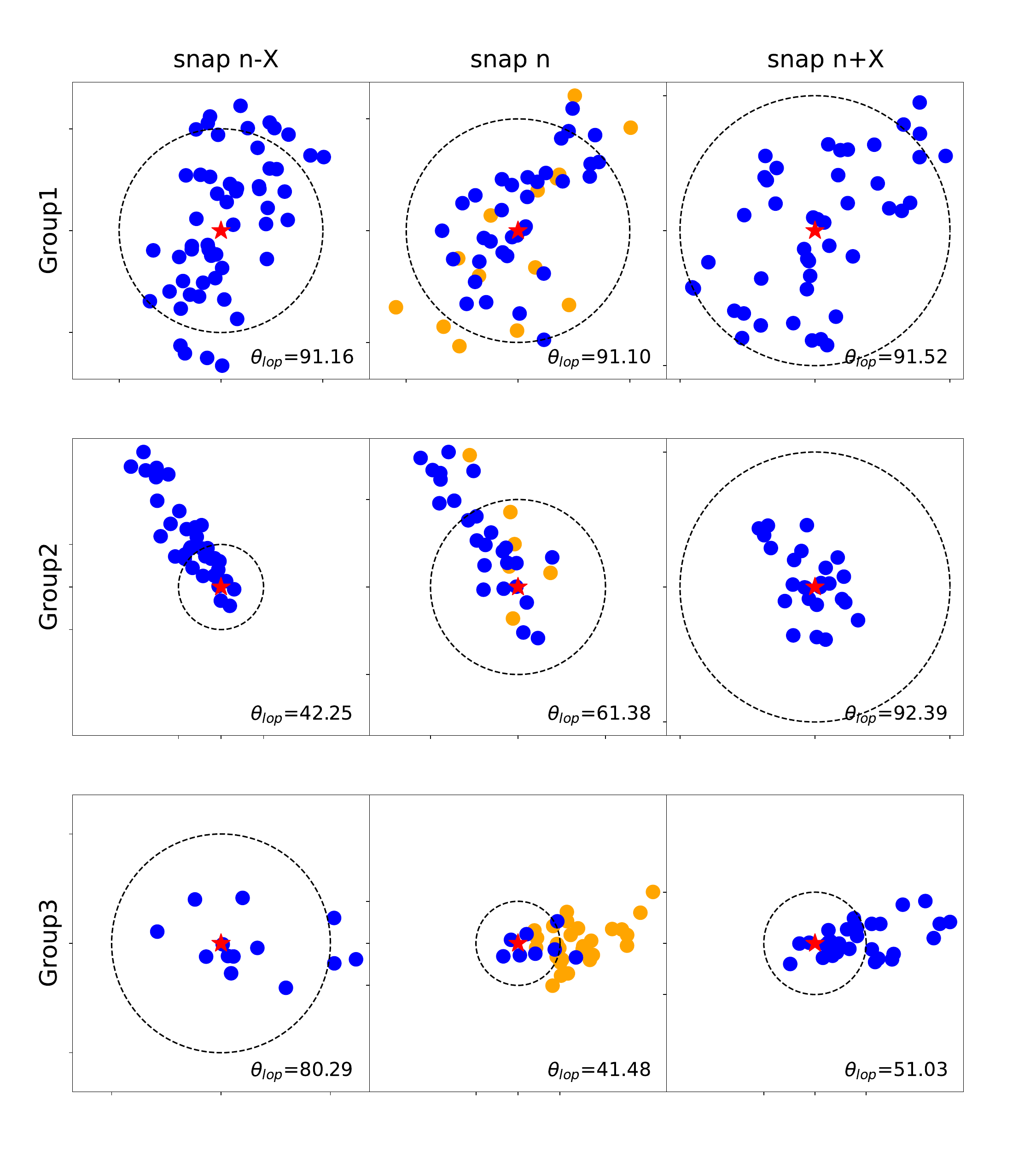}
    \caption{\textbf{The evolution of the satellite distribution of three galaxy groups is depicted from top to bottom.} The three columns from left to right represent three different snapshots corresponding to redshifts $z=0.25$, $z=0.1$, and $z=0$. The red star in the centre of each snapshot represents the central galaxy. The dashed circles indicate the viral radius at the corresponding snapshot time. The orange dots indicate satellites that were not present in the previous snapshot, while the blue dots indicate satellites that were already present in the previous snapshot. We note that the relative size of the circle does not indicate the radius size. It is used to make sure that all satellites show in the figure.}
    \label{fig:single}
\end{figure*}

Inspection of some selected examples can help to intuitively elucidate the origin of the lopsided signal and its evolutionary trajectory. In Figure~\ref{fig:single}, we show the evolution of satellite distributions for three distinct groups in three different snapshots, which we denote as snap n-X, snap n and snap n+X, corresponding to redshifts $z=0.25$, $z=0.1$ and $z=0$, respectively. The central galaxy in each group is represented by a red star. The blue dots represent the satellites that were present in the previous snapshot, while the orange dots represent satellites that have recently been accreted from the large-scale environment. The lopsided angle $\thetalop$ changes from snapshot n-X to snapshot n, showing the effect of satellite accretion from the large-scale environment and from snapshot n to snapshot n+X, corresponding to the internal evolution of satellites within the halo.

For Group~1 (top panels), the lopsided angle $\thetalop$ remains unchanged throughout the three snapshots, since the system had a large number of satellites at snapshot n-X, and the relatively small number of satellites that were accreted at snapshot n was insufficient to modify the original lopsided signal.

For Group~2 (middle panels), the satellite distribution was lopsided with $\thetalop\sim42^{\circ}$ in the snapshot n-X. The host accreted only six satellites (orange dots) in snapshot n from the large-scale environment, but the internal evolution of the existing satellites (blue dots) caused the lopsidedness to increase to $\sim61^{\circ}$. From snapshot n to snapshot n+X, satellites became almost randomly distributed ($\thetalop\sim92^{\circ}$), mainly due to the effects of internal evolution. 

In snapshot n-X, Group~3 (bottom panels) had a relatively small number of satellite galaxies, and their distribution was close to random. By snapshot n, the host had accreted more satellites from the large-scale environment, although the original distribution of the existing satellites was closer to the inner region and more randomly distributed. Most newly accreted satellites cluster in the same direction in the outer region, thus increasing the lopsided signal. At snapshot n+X, more satellites moved to the centre, causing the angle of lopsidedness to slightly increase.

We can deduce that the lopsidedness of the satellite distribution is mainly caused by two factors. The first is the satellites that fall into the dark-matter halo from the external environment, which is especially important, since most of them are accreted from a single direction, thus intensifying the signal. The second is the internal evolution of the satellites within the host dark matter halo, which causes them to move closer to the centre and randomise their distribution, thus weakening the signal.

\begin{figure}
\includegraphics[width=\columnwidth]{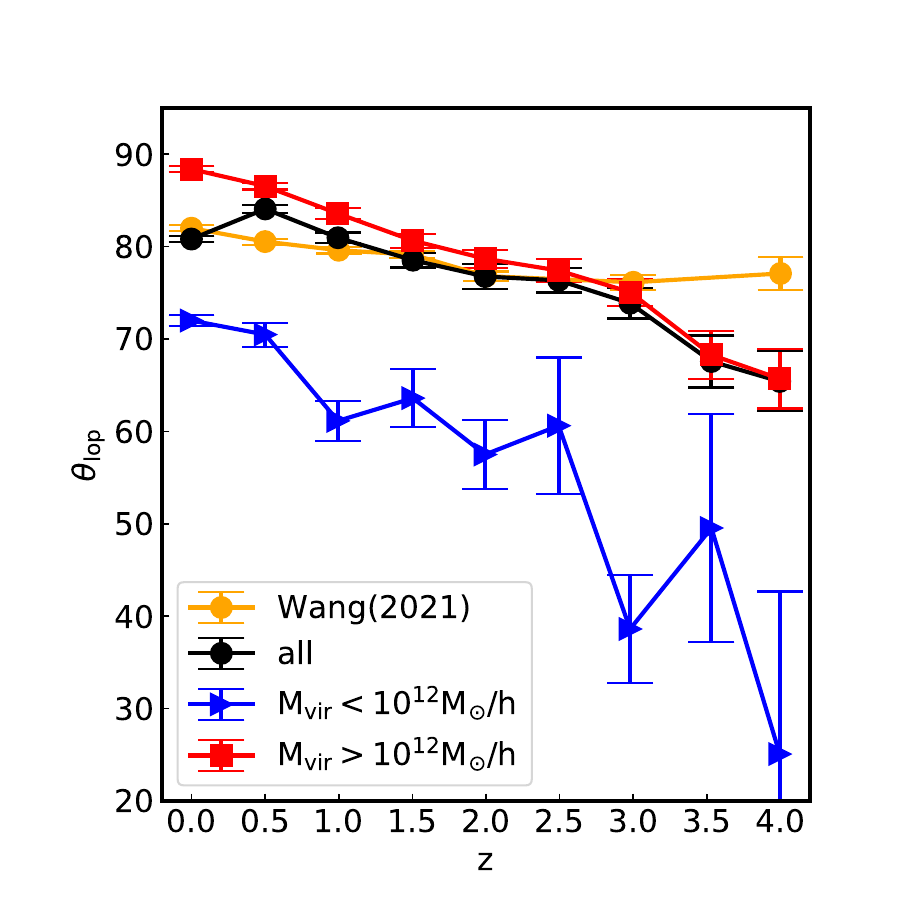}
\caption{Relation between the lopsided angle $\theta_{\rm lop}$ and redshift. The solid black line with circles is for the whole sample. The blue and red lines represent the low-mass and high-mass subsamples. The LSD shows the same weakening trend in both subsamples. The sudden decrease of the lopsided angle from $z=0.5$ to $z=0$ can be attributed to the rapidly increasing contribution of the low-mass subsample. The orange line shows result from \protect\cite{2021ApJ...914...78W}. The error bars are determined using bootstrap method.}
    \label{fig:lop_redshift}
\end{figure}

Investigating the evolution of LSD signals is a direct way to explore their origin. \cite{2021ApJ...914...78W} demonstrated an slightly increase of the lopsided angle $\theta_{\rm lop}$ (a decrease in the lopsidedness, which can be seen in the orange line of Figure~\ref{fig:lop_redshift}) as the universe evolves by studying isolated systems and calculating the lopsided angles at various redshifts without tracing the systems back in time. In this work, we trace the main progenitors of all target systems at $z=0$ back to high redshifts until the satellite number is less than 2. At $z=0$, the sample is divided into two subsamples with a halo mass cutoff of $10^{12} \msunh$.

In Figure~\ref{fig:lop_redshift}, we show the evolution of the lopsided angle $\theta_{\rm lop}$ of the main progenitors. The lopsided angle is given by the median value of all systems in each bin, and the bootstrap error is provided. For the entire sample, represented by the circles with the black line, we find that the lopsided angle $\theta_{\rm lop}$ increases from $z=4$ to $z=0$, indicating that the strength of the lopsidedness decreases from high to low redshifts. This evolutionary trend is in agreement with \cite{2021ApJ...914...78W}, although the value of $\theta_{\rm lop}$ at a given redshift is slightly different. 
The difference can be explained as follows. The criteria used by \cite{2021ApJ...914...78W} to select isolated systems remain unchanged regardless of redshift, leading to discrepancies between their sample and ours. At higher redshifts, particularly for $z>3.5$, the sample size is drastically reduced (as seen in Figure 7-f of \cite{2021ApJ...914...78W}) and the lopsided signal is driven mainly by massive groups, thus weakening the lopsided signal (with larger $\theta_{\rm lop}$).
The red and blue lines represent the low- and high-mass subsamples, respectively. The evolutionary trend of the low-mass and high-mass subsamples is comparable to that of the entire sample. However, the $\theta_{\rm lop}$ of the low-mass subsample is consistently lower (with more lopsidedness) than that of the high-mass subsample. For the entire sample, there is a sudden decrease in $\theta_{\rm lop}$ from $z=0.5$ to $z=0$. We note that this does not mean a physical increase in LSD. This is caused by the high fraction contributed by the low-mass subsample. Considering the mass limit of the satellites at $z=0$ and the fact that each progenitor must contain at least two satellites, the fraction of satellites by number decreases more rapidly than the high-mass groups.

\begin{figure*}
    \includegraphics[width=\textwidth]{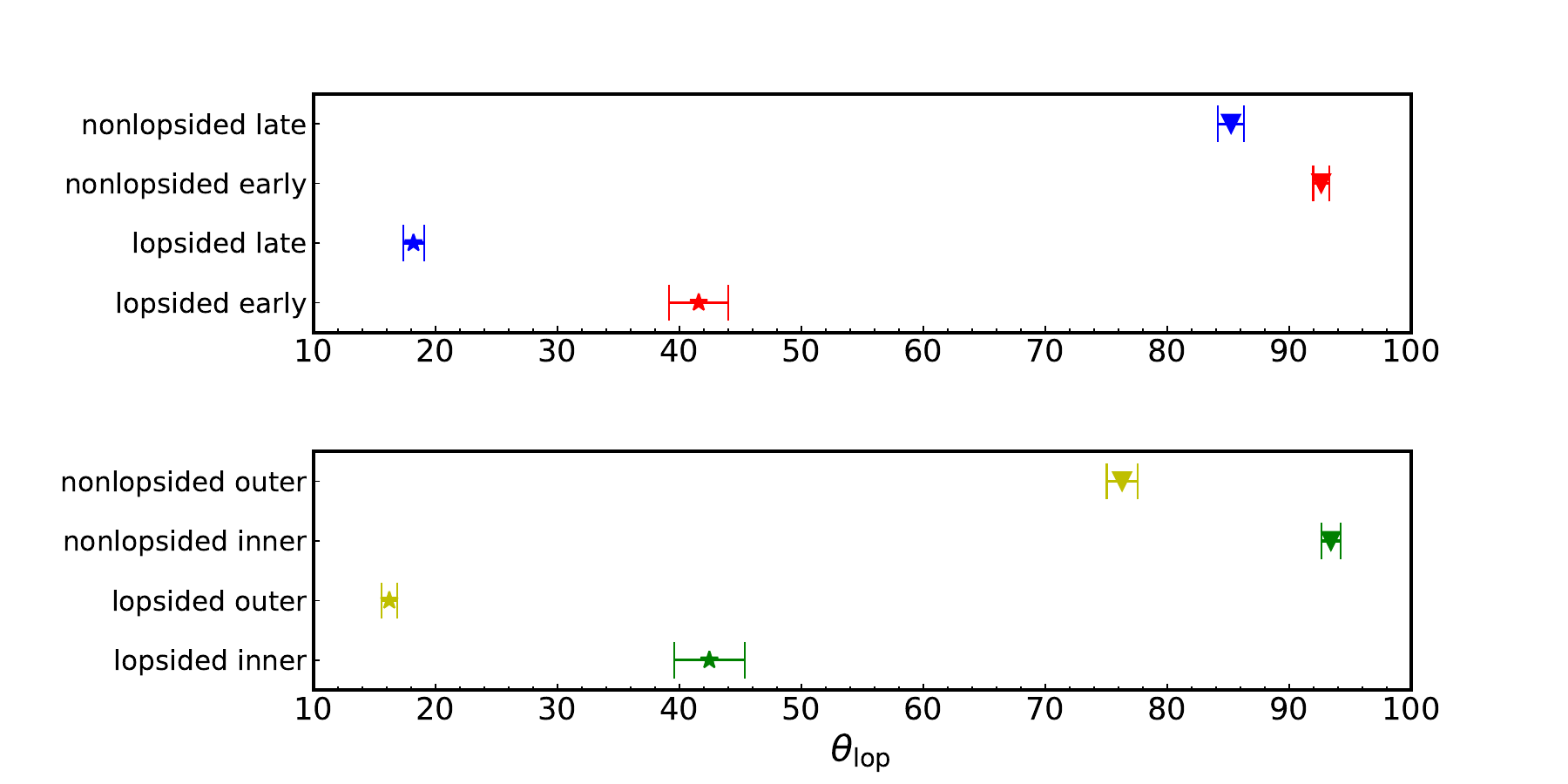}
\caption{A comparison is conducted between the lopsided angle, $\rm \theta_{lop}$, of lopsided and nonlopsided systems in terms of their satellite infall time (upper panel) and radial distribution (lower panel). The two systems are distinguished by their significance, with lopsided systems having a significance greater than $3\sigma$ and nonlopsided systems having a significance less than $1\sigma$. The error bars are determined by adding in quadrature the errors using the bootstrap method and the uncertainties from 1000 random projection directions.  }
\label{fig:inout_earlylate}
\end{figure*}

\begin{table*}
%\resizebox{\textwidth}{!}{%
\caption{The number of groups, the lopsided angle and the corresponding error for each subsample shown in Figure~\ref{fig:inout_earlylate}.}
\begin{tabular}{clcccclcccc}
\hline
                   &  & \multicolumn{4}{c}{nonlopsided} &  & \multicolumn{4}{c}{lopsided}  \\ \hline
                   &  & early  & late   & inner & outer &  & early & late  & inner & outer \\ \hline
Num                &  & 3597   & 3106   & 3340  & 2433  &  & 567   & 960   & 458   & 1079  \\
$\rm \theta_{lop}$ &  & 92.62  & 85.22  & 93.41 & 76.29 &  & 41.58 & 18.19 & 42.45 & 16.20 \\
error              &  & 0.64   & 1.13   & 0.80  & 1.23  &  & 2.52  & 0.83  & 2.94  & 0.65  \\ \hline 
\end{tabular}%
%}

\label{fig:inout_earlylate_table}
\end{table*}

\begin{figure*}
    \centering
    \includegraphics[scale=0.333]{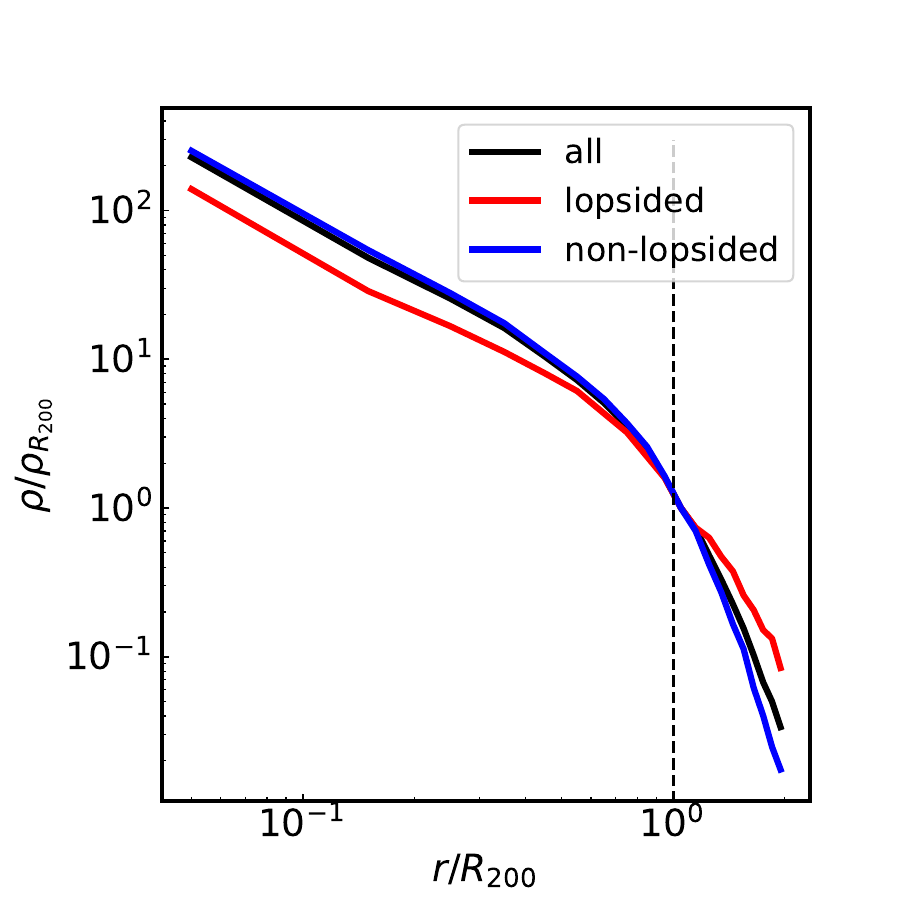}
    \hspace{2pt}
    \includegraphics[scale=0.333]{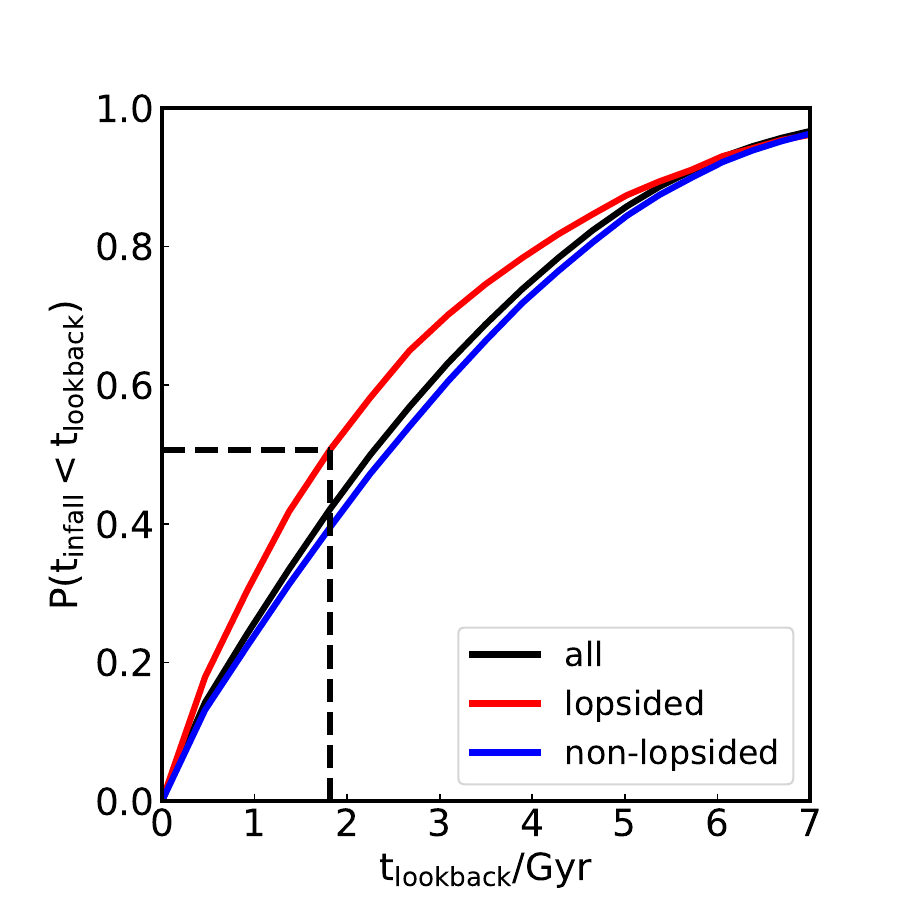}
    \hspace{2pt}
    \includegraphics[scale=0.333]{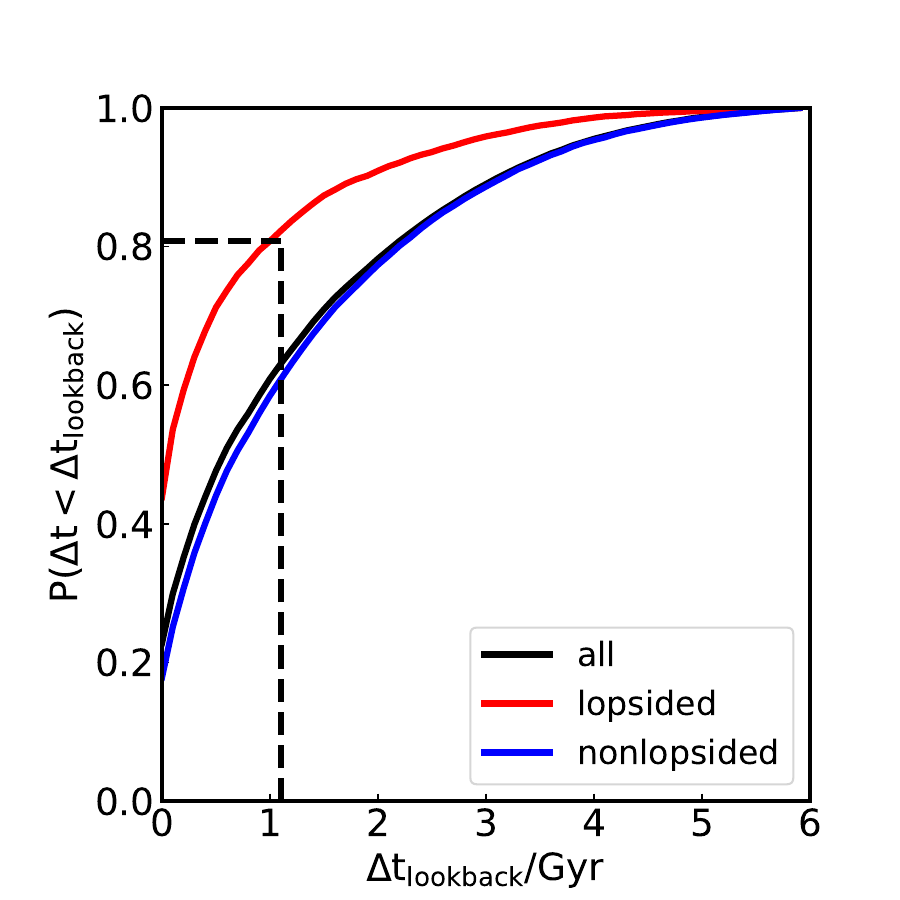}
\caption{The left panel displays the trend of the rescaled number density with respect to the radial distance from the halo centre. The middle panel shows the cumulative distribution of the time when the satellites fell into the host. The right panel illustrates the distribution of the time interval $\Delta t$ of satellites that have been accreted by the host. The solid red and blue lines in each panel represent the lopsided and nonlopsided systems, respectively, while the black dashed line represents the entire sample.}
\label{fig:comparison}
\end{figure*}

\subsection{Lopsided signal with radial distribution and infall time}
\label{subsec:t and R}

Following our previous conjecture that the LSD is dominated by external accretion from the large-scale environment and internal evolution within the halo, the spatial distribution and infall time of the satellites are therefore essential for understanding the evolution of lopsided signals. The satellite infall time, $t_{\rm infall}$, is defined as follows: if at snapshot $n$ the satellite was an independent FOF halo, while in the next snapshot $n+1$ it was a part of a larger halo but still a subhalo within the host, we use the corresponding epoch of snapshot $n$ as the infall time. The high time resolution of the outputs of MTNG makes this definition feasible. We divide the entire sample into two subsamples according to significance (as defined in Equation~\ref{equ:sig}). Systems with a significance less than $1\sigma$ ($\sim 40\%$ of the total) are labelled as \emph{not lopsided} and systems with a significance greater than $3\sigma$ ($\sim 5\%$ of the total) are labelled as \emph{lopsided} subsample. Figure~\ref{fig:inout_earlylate} shows a direct comparison of the lopsided signal, $\theta_{\rm lop}$, between the subsamples considering the infall time (upper panel) and the radial distribution (lower panel) of the satellite galaxies. The corresponding lopisded angles and errors are listed in Table~\ref{fig:inout_earlylate_table}.  

In the upper panel, we investigate the effect of the satellite infall time on lopsidedness by selecting the 30\% earliest and 30\% latest accreted satellites as early-infall and late-infall satellites, respectively, from both lopsided and non-lopsided systems. We compare the median lopsided angles for different subsamples. Two types of errors are considered for each subsample. The first type is the measurement error caused by the limited sample size, which is also listed in Table~\ref{fig:inout_earlylate_table}. This error is estimated using the bootstrap method, similar to Figure~\ref{fig:lop_redshift}. The second source of error arises from the our selected projection direction. Therefore, we generate 1000 random sightlines and estimate the error from the different median $\theta_{\rm lop}$ measurements of the 1000 sightlines (the typical error is only about $0.3^{\circ}$). The two types of errors are combined in quadrature to obtain the final errors. The lopsided angle of the early-infall satellites in lopsided systems is $41.58^{\circ}$ (red star), while the late-infall satellites have a lopsided angle of $18.18^{\circ}$ (blue star). For \emph{non-lopsided} systems, the lopsided angles of early and late infall satellites are $92.62^{\circ}$ (red triangle) and $85.22^{\circ}$ (blue triangle), respectively. The median value reveals apparent differences between early and late infall satellites, especially in lopsided systems.

In the lower panel of Figure~\ref{fig:inout_earlylate}, we divide the satellites into two groups based on their distances to the central galaxies, inner and outer satellites. We select satellites with the smallest 30\% (inner) and largest 30\% (outer) distances to their central galaxies. We then calculate the lopsided angles of these two groups for both lopsided and non-lopsided systems. For lopsided systems, the inner and outer satellites have $\theta_{\rm lop}=42.45^{\circ}$ (green star) and $\theta_{\rm lop}=16.21^{\circ}$ (yellow star), respectively. For non-lopsided systems, the corresponding values of $\theta_{\rm lop}$ are $93.41^{\circ}$ (green triangle) for the inner regions and $76.30^{\circ}$ (yellow triangle) for the outer regions. When comparing the median values between inner satellite samples and outer satellite samples, we find that the early and inner satellites have larger lopsided angles (i.e.~less lopsidedness) than the late and outer counterparts. The hierarchical galaxy formation model suggests that early-accreted satellites are mainly located in the inner part of the dark-matter halo, whereas late-accreted satellites are usually situated in the outer regions.

We show in the left panel of Figure~\ref{fig:comparison} the radial number density distributions which are rescaled by the number density $\rho$ at $R_{200}$ of satellite galaxies for lopsided (red line) and nonlopsided (blue line) systems, as well as the entire sample (black line) as a function of the relative distance to the group centres ($r/R_{200}$), where $R_{200}$ is the halo virial radius. Most of the satellite galaxies are located within $2R_{200}$. The two subsamples show considerable differences. Within $R_{200}$, the satellite number density of the \emph{nonlopsided} subsample is relatively high, while for a radius larger than $R_{200}$, the satellite number density of the \emph{lopsided} subsample is relatively high.

In the middle panel, we show the distribution of the infall time $t_{\rm infall}$ (defined in Section~\ref{subsec:t and R}) of the entire sample and the two subsamples. The right panel shows the time interval $\Delta t$ between the infall of each satellite and the previous one. We found that for \emph{lopsided} systems, nearly half of the satellites are accreted within the same time ($\Delta t=0$ but not in the same single snapshot) and 80\% within 1 Gyr (solid red line in the middle panel). For \emph{nonlopsided} systems, this fraction of $\Delta t=0$ is reduced to around 20\% (solid blue line in the middle panel). In the case of \emph{lopsided} systems, 80\% of the satellites are successively accreted in a short period of 1 Gyr (red solid line in the middle panel); however, \emph{nonlopsided} systems need 2.5 Gyr (blue solid line in the right panel) to reach that fraction.

In summary, we found that for \emph{lopsided} systems, the majority of their satellites were accreted recently over a relatively short time frame and are located in the outer parts of halos. Conversely, for \emph{nonlopsided} systems, the majority of their satellites were obtained in the past over a relatively extended period and they are situated in the inner areas of halos.

The two types of system appear to be discrepant, which results in different strengths of the lopsided signal. This could imply that the lopsided signal originates from large scales when satellites are accreted from the environment \citep{2015ApJ...813....6K,2015MNRAS.452.1052L}. They bring anisotropies from the large scale structure, but it usually decreases when moving towards the centre of the halo due to non-linear evolution within the halos. One may wonder how long this kind of memory lasts after a satellite has been accreted. To understand this, we examine the relationship between the survival time $\Delta {t}$ of satellite galaxies after accretion and the strength of their lopsided signal $\rm \theta_{lop}$ at a given redshift $z$. Survival time is defined as $\Delta t = t_{\rm infall} - t_z$, where $t_{\rm infall}$ is the infall time of a given satellite and $t_z$ is the time of a given redshift. Large $\Delta {t}$ indicates that the satellites were accreted early and evolved for a long time within the host halo.

\begin{figure}
\includegraphics[width=\columnwidth]{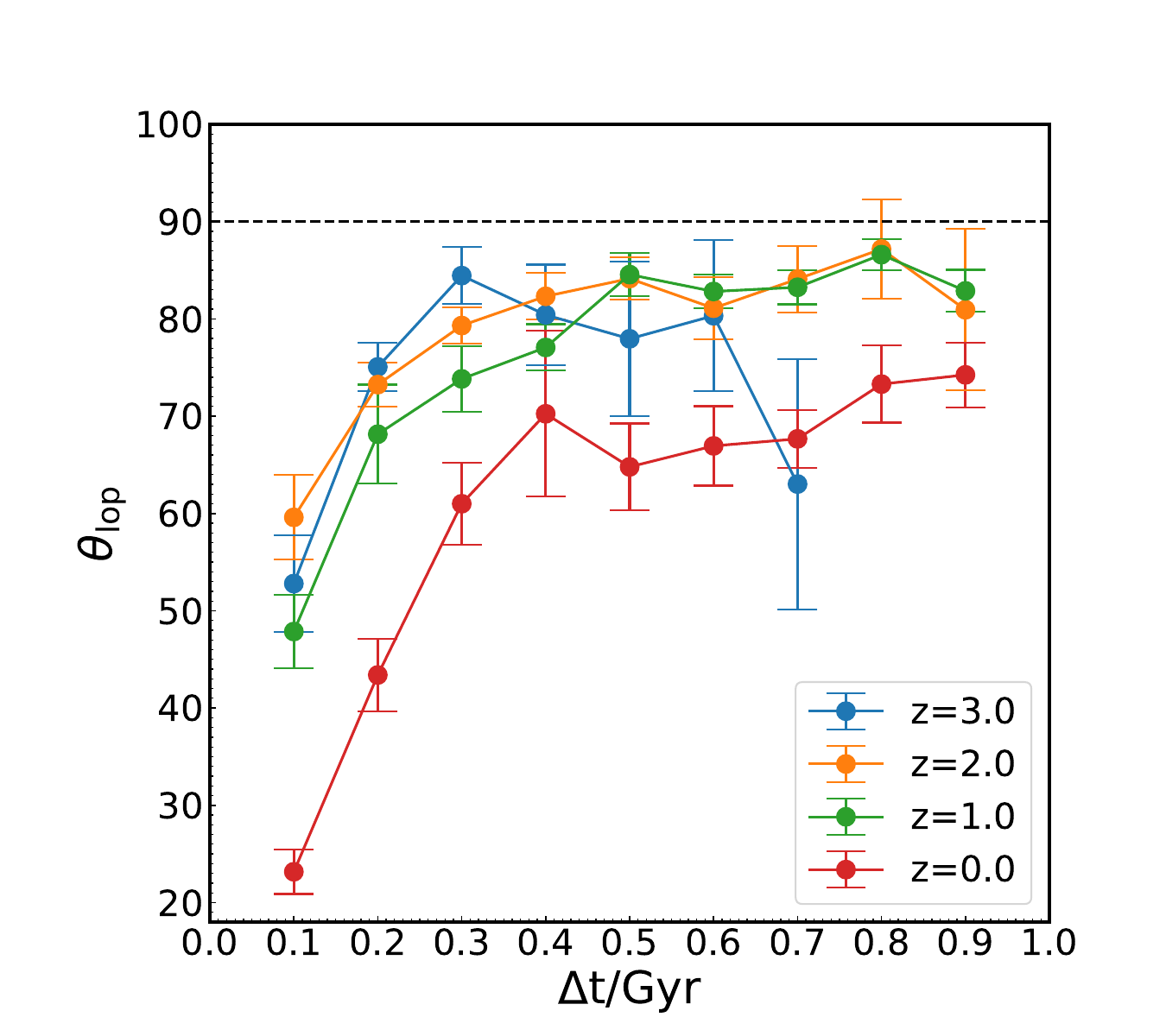}
\caption{The correlation between satellite survival time $\Delta t$ and the strength of their lopsided signal $\rm \theta_{lop}$. Satellite survival time $\Delta t$ is defined as $\Delta t = t_{\rm infall} - t_z$, where $t_{\rm infall}$ is the infall time of a given satellite and $t_z$ is the time of a given redshift $z$ (labelled). The black dashed line shows the lopsided angle of randomly distributed satellites. The error bars are determined using the bootstrap method.}
    \label{fig:infallmedian}
\end{figure}

Figure~\ref{fig:infallmedian} shows the relation between $\Delta t$ and $\theta_{\rm  lop}$. We can see a clear trend that $\theta_{\rm  lop}$ increases first with growing $\rm \Delta t$ and then becomes independent of $\Delta t$.
We find that at each $z$ (except for $z=0$ shown in the solid red line), the lopsided angle is close to $\rm 90^{\circ}$ for systems with $\Delta t \textgreater 0.3\, {\rm Gyr}$, which implies that the lopsided signal brought from infalling satellites does not survive for a long time. This trend suggests a gradual weakening of the lopsided signal introduced by infalling satellite subhalos, which eventually leads to its near disappearance after a certain period. The profiles of $z=1$, $2$, and $3$ are similar, while the profile of $z=0$ is obviously lower. This can be attributed to the large number of low-mass systems that we mentioned in Section~\ref{subsec:evolution}. The lopsided angle approaches $\rm 90^{\circ}$ at $\Delta t>1.0\, {\rm Gyr}$ and tends to become a flat profile for larger $\Delta t$, which is not shown here.

\subsection{Satellite accretion along large scale structure}
\label{subsec:largescale}

In Section~\ref{subsec:t and R}, we show that the lopsided signal is mainly due to satellites that have recently been accreted and to those that are located in the outer regions of halos. It is also observed that for lopsided systems most satellites are accreted for a short period. However, if the satellites were not accreted from a particular direction, i.e.~uniform accretion, the expected satellite distribution would still be random. \cite{2018MNRAS.476.1796S} proposed that satellites of MW-like galaxies were obtained in the group from the closest abundant filament. It has also been suggested that the subhalos were accreted along the filament direction \citep{2015ApJ...813....6K,2015ApJ...807...37S,2015MNRAS.452.1052L,2015MNRAS.450.2727T}.  Combining the fact that the lopsided signal comes from short-time satellite galaxy accretion events and that satellites fell preferentially from the filaments, we can conclude that the lopsided signal should be strongly correlated with the large-scale structure.

We can verify this by looking at the pattern of satellite accretion in relation to the large-scale structure. To do this, we first use the Cloud-in-Cell algorithm with $1024^3$ cells to generate the matter density field $\rho$. We then solve the Poisson equation ($\nabla^{2}\phi = 4\pi G\rho$) to obtain the gravitational potential $\phi$. The tidal tensor is derived from the Hessian matrix of $\phi$ as $T_{ij} = \partial_{i}\partial_{j}\phi$ at the position of the targeted system \citep{2020NewA...8001405W}. The three eigenvalues (ordered as $\lambda_1\textgreater\lambda_2\textgreater\lambda_3$) of the tidal tensor indicate the strength of the gravitational potential, and the three eigenvectors ($e_1$, $e_2$, $e_3$) show the direction of each eigenvalue. The sign of the eigenvalue can be used to determine whether the gravitational force of the eigenvector is inward (positive) or outward (negative). This allows us to distinguish the environment of each halo through the number of positive and negative eigenvalues. For example, the eigenvector $e_{3}$ associated with the smallest eigenvalue points to the direction of the strongest expansion. Therefore, the eigenvector of a halo located in a filament (which has two positive eigenvalues and one negative eigenvalue) is aligned with the direction of the filament.

We investigate the relationship between satellite lopsided distribution and large-scale structure directions by calculating the cosine of angles between the central-satellite direction $\bm{e_{i}}$ and the eigenvector $\bm{e_{3}}$ of the smallest eigenvalue, denoted $\cos(\theta_{e}) = \bm{e_{i}}\cdot \bm{e_{3}}$. A random distribution would have an expectation value $\left<\cos(\theta_{e})\right> = 0.5$. If $\cos(\theta_{e})$ is greater than 0.5, the satellites are more likely to be aligned with the large-scale direction, and if it is less than 0.5, the satellites are more likely to be perpendicular to the large-scale direction. We examine $\cos(\theta_{e})$ both at $z=0$ and at the infall time $z=z_{\rm infall}$. For $z=z_{\rm infall}$, an alignment trend indicates that satellites are more likely to be accreted along the filament direction.

\begin{figure}
\includegraphics[width=\columnwidth]{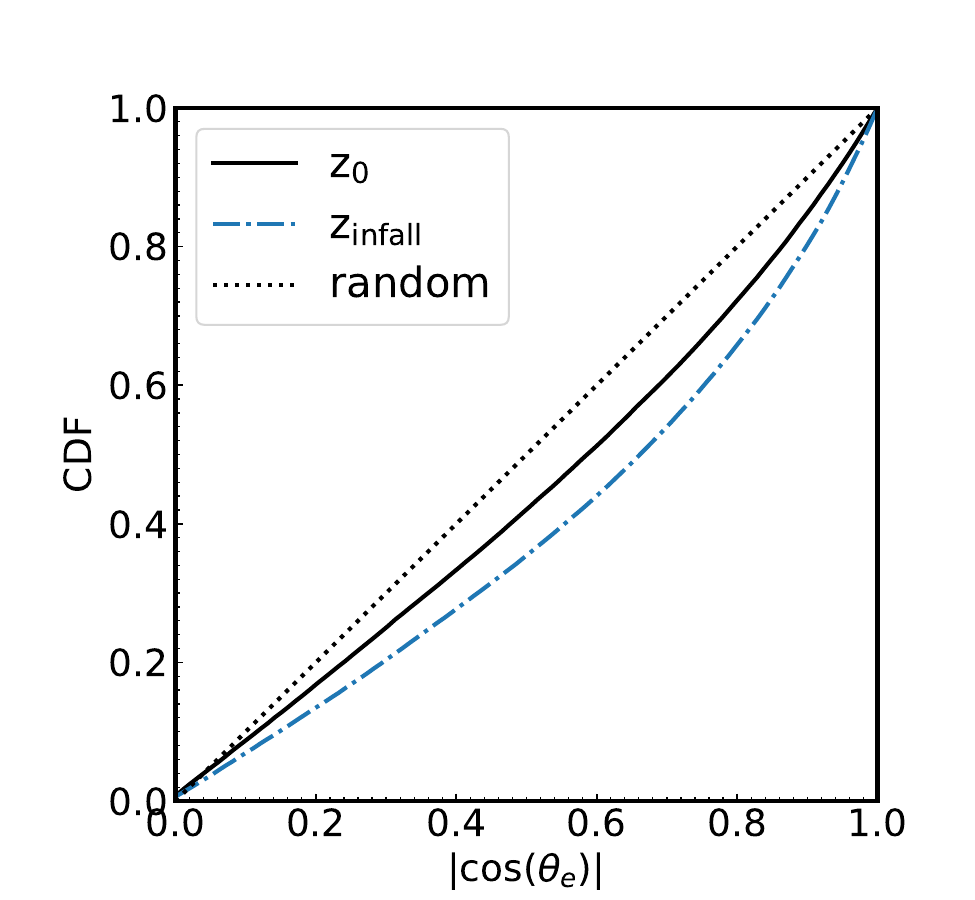}
    \caption{The cumulative distribution function of the cosine of the angle between the satellite-central vector and the primary large-scale structure direction. The black dotted line correponds to a uniform (isotropic) satellite accretion relative to the large-scale structure. The black solid line is for the satellites' position at the present time ($z=0$), while the blue dash-dotted line represents the satellites' position at the time of accretion.}
    \label{fig:largescale}
\end{figure}

In Figure~\ref{fig:largescale}, we show the cumulative distribution of the cosine of the angle between the satellite-central vector and the direction of the large-scale structure. It is clear that satellites at $z=0$ (solid black line) or $z=z_{\rm infall}$ (blue dashed dotted line) are more likely to be orientated in the direction of the large-scale structure. Moreover, there is a noticeable gap between the distributions for $z=0$ and $z=z_{\rm infall}$, indicating that the alignment at the time of accretion is stronger than at $z=0$. This implies that the satellites were accreted along the direction of the large-scale structure, which contributes to the generation of lopsided signals. Subsequently, the internal evolution effect within the host dark matter halos causes the satellites to lose their correlation with the directions of the large-scale structure, resulting in a weakening of the alignment.

%%%%%%%%%%%%%%%%%%%%%%%%%%%%%%%%%%%%%%%%%%%%%%%%%%%%%
%%%%%%%%%%%%% 
%   con & dis
%%%%%%%%%%%%% 
\section{Summary and Discussion}\label{sec:sum_dis}

Observations and simulations have both revealed an asymmetric distribution of satellites around paired galaxies \citep{2016ApJ...830..121L,2017ApJ...850..132P} and isolated systems \citep{2020ApJ...898L..15B,2021ApJ...914...78W}. While the origin of this lopsided distribution has been explored in paired galaxies \cite{2019MNRAS.488.3100G}, the physical origin of the same phenomenon in isolated systems is still unknown. This paper investigates the origin of the lopsided satellite distribution around isolated central galaxies, using the state-of-the-art hydrodynamical $\Lambda$CDM simulation MillenniumTNG. Our findings can be summarised as follows.

\begin{itemize}
\item We have re-examined the lopsided signal of the isolated system selected at $z=0$. Our results agree with \cite{2021ApJ...914...78W} (see Figure~\ref{fig:lopsided}).

\item The lopsided signal decreases with cosmic time, and this trend is mass-dependent.  High-mass systems experience slower fading, whereas the reduction is more rapid for low-mass systems (see Figure~\ref{fig:lop_redshift}).

\item The lopsided signal at $z=0$ is dominated by satellites located at the outer parts of halos and is also dominated by recently accreted satellites (see Figure~\ref{fig:inout_earlylate}).

\item The lopsided signal originates from the anisotropic accretion of satellite galaxies from the large-scale structure (see Figure~\ref{fig:largescale}), and after accretion, the non-linear evolution of satellites inside the dark matter halos weakens the lopsidedness (see Figures~\ref{fig:infallmedian} and \ref{fig:largescale}).
\end{itemize}

Our findings for the lopsided distribution of satellite galaxies at $z=0$ are in general agreement with the results of \cite{2020ApJ...898L..15B,2021ApJ...914...78W,2023ApJ...947...56S}. In addition, we have explored the origin of this lopsidedness in isolated systems. We studied whether the lopsided satellite distribution is caused by the external large-scale structure or the internal evolution of dark matter halos. By tracing systems back to the early Universe, we examined the satellite accretion pattern with respect to the large-scale structure and satellite infall time, as well as the radial satellite distribution. We identified the generation and evolution of lopsided signals from satellites.

The large-scale structure has a significant impact on the formation of lopsided satellite distributions, as it is strongly associated with the pattern of satellite accretion, which is characterised by a preferential accretion of satellites along certain large-scale structure directions. This is in agreement with the findings of \cite{2019MNRAS.488.3100G}, who suggest that the lopsidedness of both central pairs and isolated systems is the result of the specific direction of satellite fall from the environment \citep{2014MNRAS.443.1274L,2015ApJ...813....6K}. The more and the later satellites that fall into a galaxy group, the more pronounced the lopsided signal becomes. Late-infalling satellites, which are more likely to be located in the outer parts of dark matter halos, are particularly responsible for the lopsidedness.

Once the satellites are accreted, the internal evolution within the dark matter halo and the external origin of the satellites begin to compete. The relaxation of the dark matter halo causes the satellites to become more isotropically distributed with time. The closer a satellite galaxy is to the inner regions of a halo, the longer it has experienced the influence of relaxation, making it lose more of the memory it had from the large-scale structure.

Despite residing in the inner regions of the host halo, such satellites nevertheless still exhibit some weak lopsidedness. What other factors could be suppressing the relaxation of this signal, preventing it from dissipating completely? How long does it take for the lopsidedness caused by the large scales to be completely relaxed? The source of the lopsidedness from the large-scale structure at early epochs has yet to be explored. This is why satellites are accreted into dark matter halos in groups and aligned with the large-scale structure. Further research is required to construct a full model of the generation of anisotropy of the satellite distribution at high redshifts.

In addition, the lopsided distribution of satellites that we have investigated in this work, as well as in previous studies, is based on the definitions of central and satellite galaxies provided by the FOF and SUBFIND-HBT algorithms. However, bridge-like configurations sometimes caused by the FOF algorithm may affect the lopsided signal by biasing the definition of the centre of the system.  For example, in the case of two merging systems with a very similar mass of the central galaxies, the algorithm may identify them as a single system. In such a double-centred system, the algorithm will designate only one galaxy as central, and the others as satellite galaxies, potentially creating a spurious lopsided signal. This issue will be explored further in our future work.

%%%%%%%%%%%%%%%
%     Acknowledgments
%%%%%%%%%%%%%%%
\section*{Acknowledgements}
We thank the anonymous reviewr for the helpful comments that significantly improved the presentation of this paper.
We acknowledge the stimulating discussions with Prof. Noam I. Libeskind and Prof. Xi Kang. This work is supported by the National SKA Program of China (grant No. 2020SKA0110100). PW is sponsored by Shanghai Pujiang Program(No. 22PJ1415100). HG is supported by the CAS Project for Young Scientists in Basic Research (No. YSBR-092) and the science research grants from the China Manned Space Project with NO. CMS-CSST-2021-A02. SB is supported by the UK Research and Innovation (UKRI) Future Leaders Fellowship [grant number MR/V023381/1]. We acknowledge the use of the High Performance Computing Resource in the Core Facility for Advanced Research Computing at the Shanghai Astronomical Observatory. 

%%%%%%%%%%%%%%%
%     Acknowledgments
%%%%%%%%%%%%%%%
\section*{Data availability}
The MillenniumTNG simulations will be publicly available on \url{https://www.mtng-project.org} in the future. The data we use in this article will be shared upon reasonable request to the corresponding author.

%%%%%%%%%%%%%%%%%%%%%%
%    Bibliography
%%%%%%%%%%%%%%%%%%%%%%
\bibliographystyle{mnras}
\bibliography{export-bibtex}
%\bibliography{export-bibtex}

% %%%%%%%%%%%%The End%%%%%%%%%%%%%%%%%%%%%%%%%%%%%%%%%%%%%%%%%

%%%
%%%\begin{figure}[!ht]
%%%    \includegraphics[scale=0.6]{massbin.pdf}
%%%    \hspace{10pt}
%%%    \includegraphics[scale=0.6]{massnum.pdf}
%%%    \caption{Left panel: Dark matter halo mass distribution of the selected sample. The black dashed line shows the median value of the halo mass. Right panel: The number of systems for higher or lower than $10^{11.8} \msunh$ at different redshift. At $\rm z\sim 0$ the whole sample is divided into two equally-sized sub-samples according to their halo mass while at high redshift the number of low-mass systems rapidly decreased and the high-mass counterpart dominates the sample.}
%%%    \label{fig:largescale}
%%%\end{figure}
%%%\end{appendix}

% \label{lastpage}
\end{document}